\documentclass[letterpaper,preprintnumbers,prd,twocolumn,nofootinbib,nobibnotes,showpacs]{revtex4}
\usepackage{amsfonts}
\usepackage{mathrsfs}
\usepackage{epsfig}
\usepackage{graphicx}%
\usepackage{dcolumn}
\usepackage{amsmath}
\usepackage{color}
\usepackage{color}
\makeatletter
\def\btt#1{\texttt{\@backslashchar#1}}%
\DeclareRobustCommand\bblash{\btt{\@backslashchar}}%
\makeatother
\begin{document}

\title{Cyclic universe due to phantom and quintessence}
\author{Changjun Gao}\email{gaocj@bao.ac.cn} \affiliation{The National
Astronomical Observatories, Chinese Academy of  Sciences, Beijing,
100012, China} \affiliation{State Key Laboratory of Theoretical
Physics, Institute of Theoretical Physics, Chinese Academy of
Sciences, Beijing 100190, China}
\author{Youjun Lu}\email{luyj@bao.ac.cn} \affiliation{The National
Astronomical Observatories, Chinese Academy of  Sciences, Beijing,
100012, China}
\author{You-Gen Shen}
\email{ygshen@center.shao.ac.cn} \affiliation{Shanghai
Astronomical Observatory, Chinese Academy of Sciences, Shanghai
200030, China}

\date{\today}

\begin{abstract}

We explore a cyclic universe due to phantom and quintessence
fields. We find that, in every cycle of the evolution of the
universe, the phantom dominates the cosmic early history and
quintessence dominates the cosmic far future. In this model of
universe, there are infinite cycles of expansion and contraction.
Different from the inflationary universe, the corresponding cosmic
space-time is geodesically complete and quantum stable. But
similar to the Cyclic Model, the flatness problem, the horizon
problem and the large scale structure of the universe can be
explained in this cyclic universe.

\end{abstract}


\maketitle

\section{Motivation}

The theory of inflationary cosmology has become the standard paradigm
for the early universe, which was firmly established in the past
several decades.  Inflationary cosmology not only naturally solves the
flatness problem, the horizon problem and the magnetic monopole
problem, but also elegantly explains the large scale structure of the
universe.  Despite of its success, it may be not perfect yet and
suffer from some conceptual problems as follows.

First, a few common questions may be almost always asked as
follows: What is the inflaton field? How to fine-tune the scalar
potential in order that it is sufficiently flat?  How did the
universe begin and why did it start to inflate? These questions
are of great importance for the theory consummation. Second,
inflationary cosmology suffers from the trans-Planckian problem
\cite{robert:12}. In inflationary cosmology, the fluctuations
observed in the cosmic microwave background had wavelengths at the
beginning of inflation smaller than the Planck scale. It is
generally believed that the quantum gravity would become important
at sub-Planckian scales. So the quantum gravity may produce
different distribution on sub-Planckian wavelengths. This
different distribution would be inflated and generate an uncertain
correction to the predictions for the cosmic microwave background
anisotropy. Third, Borde el al \cite{borde:01} showed that the
inflationary space-time is geometrically incomplete in the past
directions of the history of the universe as the expansion rate
averaged along the geodesic is positive, i.e., $\left <H \right>
>0$. This incompleteness theorem relies on neither the energy
condition nor the Einstein's equation. In order to solve these
problems, a number of alternative scenarios, e.g., the cyclic
universe \cite{paul:02,baum:07}, the bounce universe
\cite{bounce:10} and the emergent universe \cite{ell:046, ell:047,
ell:048, ell:049, ell:0410, ell:0411} are proposed.

In the Cyclic Model \cite{paul:02}, the universe undergoes an
endless sequence of cosmic expansion and contraction. The cosmic
temperature and energy density are all finite at each transition
point from contraction to expansion or from expansion to
contraction.  The Cyclic Model can solve not only the flatness
problem, the horizon problem, the large scale structure problem,
but also the Big-Bang singularity problem, the current cosmic
accelerating problem and so on. Therefore, the cyclic model may
present a complete description of all phases of the cosmic
evolution.

The cyclic universe model \cite{paul:02,baum:07} is originally
constructed in the brane-world scenario. In this paper, we
alternatively construct a cyclic universe by simply introducing a
phantom matter and quintessence matter. In this cyclic universe,
there are infinite cycles of expansion and contraction. Different
from the inflationary universe, the corresponding cosmic
space-time is geometrically complete and quantum stable. Similar
to the Cyclic Model, the flatness problem, the horizon problem and
the large scale structure of the universe can be explained in this
cyclic universe.

This paper is organized as follows. In
section~\ref{sec:chameleon}, we present the energy density for the
phantom and quintessence, respectively. Then, by taking into
account both the relativistic  matter and dust matter,  we
construct a cyclic model for the universe and investigate its
cyclic evolution
in section~\ref{sec:model}. 
The total cosmic energy density is zero at the transition points,
either from expansion to contraction or from contraction to
expansion. Therefore, the Big-bang and Big-crunch singularity
problems are automatically vanishing. In section~\ref{sec:geo}, we
demonstrate that the cosmic space-time is geodesically complete
for both time-like and null geodesics. The particle and the event
horizon are all infinite in this space-time. So the horizon
problem is solved. In section~\ref{sec:entr}, we discuss the
variation of entropy of the cyclic universe and the violation of
second law of thermodynamics due to the presence of phantom
matter. In section~\ref{sec:quant}, using the Wheeler-DeWitt
equation, we find that the {system} is quantum stable although a
negative scalar potential is present. In
section~\ref{sec:scaleinvariant}, we verify that the primordial
scale-invariant power spectrum problem could be generated in this
cyclic model. Finally, conclusions and discussions are given in
section~\ref{sec:conclusion}.  Throughout this paper, we adopt the
system of units in which $G=c=\hbar=1$ and the metric signature
$(-,\ +,\ +,\ +)$.

\section{phantom and qunitessence}\label{sec:chameleon}
Phantom matter is introduced into in the study of cosmic evolution
by Caldwell in Ref.~\cite{caldwell:99}. The Lagrangian of phantom
matter takes the form
\begin{eqnarray}
\mathscr{L}_{\phi}=-\frac{1}{2}\nabla_{\mu}\phi\nabla^{\mu}\phi-V\left(\phi\right)\;,
\end{eqnarray}
with $\phi$ the phantom field and $V(\phi)$ the phantom potential.
For our purpose, we shall simply take a vanishing potential. Then
it is straightforward to show the phantom $\phi$ with the
Lagrangian
\begin{eqnarray}
\mathscr{L}_{\phi}=-\frac{1}{2}\nabla_{\mu}\phi\nabla^{\mu}\phi\;,
\end{eqnarray}
contributes the energy density
\begin{eqnarray}
\rho_{\phi}=-\frac{\rho_{\phi_{0}}}{a^6}\;,
\end{eqnarray}
in the background of spatially flat Friedmann-Robertson-Walker
Universe. Here $\rho_{\phi_{0}}$ is an integration constant which
has the physical meaning of the minus of energy density when the
cosmic scalae factor $a=1$.

Compared to the Lagrangian of quintessence
field~\cite{quintessence:88}, the sign before the kinetic term of
phantom is not positive, but negative. So the kinetic energy
density of phantom becomes negative. For simplicity, we shall
consider the quintessence field with the Lagrangian as follows
\begin{eqnarray}
\mathscr{L}_{\psi}=\frac{1}{2}\nabla_{\mu}\psi\nabla^{\mu}\psi+\frac{\rho_{\Lambda}}{3}\left[1-2\sin\left(4\sqrt{\pi}\psi\right)\right]\;,
\end{eqnarray}
which contributes the energy density
\begin{eqnarray}
\rho_{\psi}=\rho_{\Lambda}-{\rho_{\psi_{0}}}{a^2}\;.
\end{eqnarray}
Here $\rho_{\Lambda}$ and $\rho_{\psi_{0}}$ are two integration
constants. Th reason for these specific choices will be given in
the next section and we shall show that, with these terms, a
cyclic universe is constructed.

\section{Cyclic universe}\label{sec:model}

Now the Friedmann equation could be written as follows
\begin{eqnarray}\label{eq:f}
3H^2&=&8\pi\left(-\frac{\rho_{\phi_{0}}}{a^6}+\frac{\rho_{r_{0}}}{a^4}
+\frac{\rho_{m_{0}}}{a^3}+\frac{\rho_{k}}{a^2}+\rho_{\Lambda}-{\rho_{{\psi_{0}}}}{a^2}\right)\;.
\nonumber \\
\end{eqnarray}
On the right hand side of Eq.~(\ref{eq:f}), the energy densities
are from phantom field, relativistic matter, dust matter (which
includes dark matter and baryonic matter), cosmic spatial
curvature  and quintessence, respectively. The parameters
$\rho_{\phi_{0}}$, $\rho_{r_0}$, $\rho_{m0}$, $\rho_k$,
$\rho_{\Lambda}$, and $\rho_{\psi0}$ represent the present-day
energy density of each component, respectively. The parameter
$\rho_k$ can be positive, negative or $0$, and the other five
parameters are all positive.

Now we could understand why we introduce the negative scalar
potential. According to the Friedmann equation for the standard
$\Lambda \textrm{CDM}$ cosmology
\begin{equation}
3H^2 =
8\pi\left(\frac{\rho_{r_{0}}}{a^4}+\frac{\rho_{m_{0}}}{a^3}+\frac{\rho_{k}}{a^2}+\rho_{\Lambda}\right)\;,
\label{eq:fri}
\end{equation}
the evolution of the scale factor $a$ is controlled by the
relativistic matter if $a$ is sufficiently small, and $a\propto
\sqrt{t}$. At the Big-bang singularity, $t=0$, the scale factor
vanishes and the energy density is divergent. In order to erase
the Big-bang singularity and turn the universe from contraction to
expansion, namely, making the velocity of the universe $\dot{a}$
varying from negative (contracting universe) to positive
(expanding universe), it is necessary to introduce the negative
energy density. At the turning point, we should have $\dot{a}=0$.

For simplicity, the negative energy density may be assumed to scale as
$a^{-n}$ with $n>4$ such that the velocity of the universe $\dot{a}$
could cross zero at the minimum of the scale factor. We take $n=6$ in
this paper. Similarly, with the increasing of the scale factor, dark
energy dominates finally. In order to let the universe contract at
sometime in the future, it is also necessary to introduce negative
energy density that dominates at large $a$, and this density may be
assumed to scale as $a^{m}$ with $m>0$ such that the velocity of the
universe $\dot{a}$ could cross zero at the maximum of the scale
factor. In this study, we simply take $m=2$.

Given the Friedmann equation Eq.~(\ref{eq:f}), and the equation of
conservation for energy, we derive the acceleration equation
\begin{equation}
6\frac{\ddot{a}}{a} = -8\pi\left(-4\frac{\rho_{\phi_{0}}}{a^6}+2\frac{\rho_{r_{0}}}{a^4}+\frac{\rho_{m_{0}}}{a^3}-2\rho_{\Lambda}+4{\rho_{{\psi_{0}}}}{a^2}\right)\;.
\label{eq:acc}
\end{equation}

For the present-day universe, we have the Friedmann equation
\begin{eqnarray}\label{eq:pf}
3H_0^2&=&8\pi\rho_0\;,
\end{eqnarray}
where $H_0$ and $\rho_0$ are the present-day Hubble parameter and
total cosmic energy density.

Dividing the acceleration equation Eq.~(\ref{eq:acc}) by
Eq.~(\ref{eq:pf}), we obtain
\begin{equation}
\frac{2}{H_0^2}\cdot\frac{\ddot{a}}{a} = \frac{4\Omega_{\phi_{0}}}{a^6}-\frac{2\Omega_{r_{0}}}{a^4}
-\frac{\Omega_{m_0}}{a^3}+2\Omega_{\Lambda}-4{\Omega_{{\psi_{0}}}}{a^2}\;,
\label{eq:main}
\end{equation}
where the dimensionless constants (the ratio of each component in
the present-day cosmic energy density) are defined as
\begin{eqnarray}
&&\Omega_{\phi_{0}}\equiv\frac{\rho_{\phi_0}}{\rho_0}\;,\ \ \
\Omega_{r_{0}}\equiv\frac{\rho_{r_0}}{\rho_0}\;,\ \ \
\Omega_{m_{0}}\equiv\frac{\rho_{m_0}}{\rho_0}\;,\ \ \nonumber\\&&
\Omega_{\Lambda}\equiv\frac{\rho_{\Lambda}}{\rho_0}\;,\ \ \
\Omega_{\psi_{0}}\equiv\frac{\rho_{\psi_0}}{\rho_0}\;,\ \ \
\Omega_{k}\equiv\frac{\rho_{k}}{\rho_0}\;.\ \ \
\label{eq:def}
\end{eqnarray}
We can let the cosmic time $t$ absorb $H_0$ in
Eq.~(\ref{eq:main}). Then the unit of cosmic time is the
present-day Hubble age
\begin{eqnarray}
\tau=\frac{1}{H_0}\;.
\end{eqnarray}
Eq.(\ref{eq:main}) is reduced to
\begin{equation}
\frac{2\ddot{a}}{a} =
\frac{4\Omega_{\phi_{0}}}{a^6}-\frac{2\Omega_{r_{0}}}{a^4}
-\frac{\Omega_{m_0}}{a^3}+2\Omega_{\Lambda}-4{\Omega_{{\psi_{0}}}}{a^2}\;.
\label{eq:main9}
\end{equation}

\begin{figure}[h]
\begin{center}
\includegraphics[width=9cm]{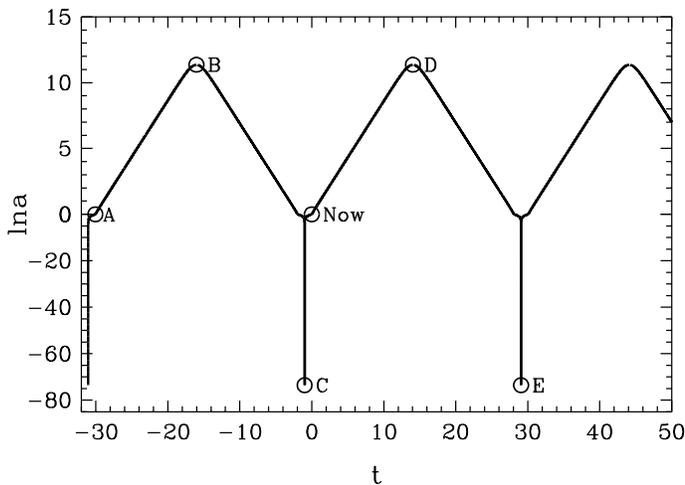}
\caption{Evolution of the scale factor $\ln{a}$ with the cosmic
time (in unit of the present-day Hubble age). }
\label{a}
\end{center}
\end{figure}

As an example, we consider the cosmology model with the following
values of parameters, $\Omega_{r_0}=8.1\cdot10^{-5}$,
$\Omega_{m_0}=0.27$, $\Omega_{\Lambda}=0.73$, and $\Omega_{k}=0$,
which are consistent with current observations \cite{sper:06}. As
for the values of $\Omega_{\phi_0}$ and $\Omega_{\psi_0}$, they
must be sufficiently small in order to be not contradict with the
well constrained standard $\Lambda \textrm{CDM}$ model. However,
with decreasing scale factor, the universe is finally dominated by
both the $\Omega_{\phi_0}$ term and the radiation. A maximum
energy density appears in this epoch. It is generally believed
that the Planck energy density,
\begin{eqnarray}\label{eq:planck}
\Omega_{\rm Planck}\simeq10^{123}\;,
\end{eqnarray}
is the highest one. So $\Omega_{\phi_0}$ is constrained to be in the
order of $\Omega_{\phi_0}\simeq 10^{-68}$. For $\Omega_{\psi_0}$, we
simply set $\Omega_{\psi_0}\simeq 10^{-10}$, which is not inconsistent
with astronomical observations.

\begin{figure}[h]
\begin{center}
\includegraphics[width=9cm]{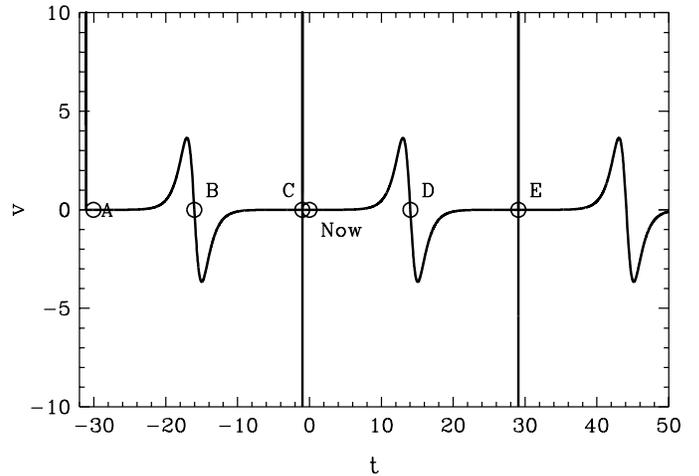}

\caption{Velocity $v\equiv10^{-4}\dot{a}$ of the universe with the
cosmic time $t$ (in unit of the present-day Hubble age). }

\label{v}
\end{center}
\end{figure}

\begin{figure}[h]
\begin{center}
\includegraphics[width=9cm]{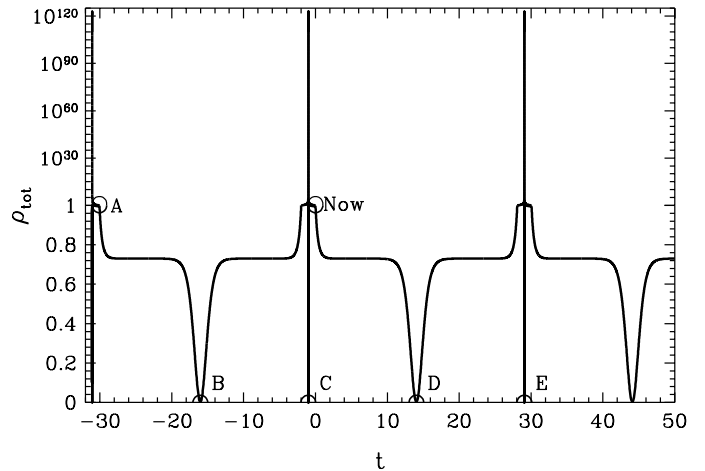}

\caption{Evolution of the total cosmic energy density ($\rho$) with
the cosmic time $t$. The unit of $t$ and $\rho$ is the present-day
Hubble age and total cosmic energy density, respectively.}

\label{p}
\end{center}
\end{figure}

Figures~\ref{a} and \ref{v} show the evolution of the scale factor
$\ln a$ and the velocity $v \equiv10^{-4}\dot{a}$ of the universe with
respect to the cosmic time $t$. Apparently there are infinite cycles,
expansion and contraction, in the evolution of the universe. The
present-day universe locates at $\ln a=0$ as we have set $a=1$ for the
present-day universe.  The minimum and the maximum of the scale factor
are $a_{\rm min}\simeq \left(\Omega_{\phi0}/\Omega_{r0}
\right)^{1/2} \sim e^{-73}$ and $a_{\rm max} \simeq
\left(\Omega_{\psi0}/\Omega_{\Lambda}\right)^{1/2}\sim e^{11}$,
respectively (see Fig.~\ref{a}). Note here that what characters
the physical size of the universe is not the scale factor, but the
apparent horizon, event horizon and particle horizon. We shall come
back to this point in the next section.

Fig.~\ref{p} shows the evolution of total cosmic energy density
$\rho$ with respect to the cosmic time. The maximum and minimum of
the energy density are the Planck density and zero, respectively.
As seen from Figure~\ref{p}, the universe contracts from $B$ to
$C$ and expands from $C$ to $D$ with the increasing of time. The
contracting and expanding phase constitute a complete cycle of
evolution. In order to understand the evolution in detail, let's
start from point $A$, which represents the time one cycle period
before the present day (point Now in Fig.~\ref{p}). In the first
several Hubble ages (a plateau), the universe is dominated by the
term $\Omega_{\Lambda}$. Then the term $-\Omega_{\psi_0}a^2$ grows
and the total cosmic energy density evolves as
$\rho\simeq\Omega_{\Lambda}-\Omega_{\psi_0}a^2$. The energy
density decreases with the expansion of the universe. When the
density vanishes, we find from the Friedmann equation
Eq.~(\ref{eq:f}) that the velocity $\dot{a}$ of the universe also
vanishes but the acceleration of the universe $\ddot{a}$ becomes
negative (from Eq.~(\ref{eq:main9})). So after crossing point $B$,
the universe evolves into contracting phase. $B$ (and $D$)
corresponds to the maximum of the scale factor as clearly shown in
Fig.~\ref{a}.

In the contracting phase (from $B$ to $C$), the universe is first
dominated by $\Omega_{\Lambda}-\Omega_{\psi_0}a^2$, and then by
$\Omega_{\Lambda}$ (the plateau), by the matter, by the radiation,
and by radiation and term $\Omega_{\phi_0}$ in a sequence. The
maximum of energy density appears at $a\simeq \frac{\sqrt{6}}{2}
a_{\rm min}$. With the increasing of term $\Omega_{\phi_0}$, the
comic density decreases sharply to $0$ at $a = a_{\rm min}$. At
the minimum of the scale factor, both the cosmic density and the
cosmic velocity vanishes (point $C$). After point $C$, the
universe goes into the expanding phase. {At the turning point $C$,
we have $\dot{a}=0$ and $\ddot{a}>0$ which follows from
Eq.~\ref{eq:f} and Eq.~\ref{eq:main9}.}

In the expanding phase (from $C$ to $D$), the universe is firstly
dominated by both the chameleon scalar and the radiation,
$\rho\simeq\Omega_{r_0}/a^4-\Omega_{\phi_0}/a^6$. Since the
density of chameleon scalar decreases quickly with the expansion
of the universe, the cosmic energy density grows significantly.
Once the cosmic energy density crosses the maximum, the universe
evolves as the standard $\Lambda \textrm{CDM}$ model (before
arriving at point $D$). The present-day universe locates at
$\rho=1$ (one unit of present-day cosmic energy density,
$\rho_0$). It is apparent the age of the universe (from the
maximum energy density to Now) is within one Hubble age which is
consistent with observations.

\section{geodesic complete space-time and
horizons}\label{sec:geo}

\subsection{geodesic complete space-time}

In this sub-section, we show that the space-time of the cyclic
universe constructed in section~\ref{sec:model} is complete for
both null and time-like geodesics. Therefore, there is no
singularity in this space-time. To this end, let's first derive
the equation of motion for a photon and a massive particle. The
metric of spatially-flat Friedmann-Robertson-Walker Universe is
give by
\begin{eqnarray}
ds^2=dt^2-a\left(t\right)^2\left(dx^2+dy^2+dz^2\right)\;.
\end{eqnarray}
Then the Lagrangian for a photon and a massive particle
propagating in this space-time is given by \cite{chand:83}
\begin{eqnarray}
\mathscr{L}=\frac{1}{2}\left[\dot{t}^2-a^2\left(\dot{x}^2+\dot{y}^2+\dot{z}^2\right)\right]\equiv\epsilon\;,
\label{eq:laggeo}
\end{eqnarray}
where dot denotes the derivative with respect to the affine
parameter $\tau$, and $\tau$ is exactly the proper time of a massive
particle. We have $\epsilon=0$ and $\epsilon=1/2$ for the photon and
massive particle (with mass of unit one), respectively
\cite{chand:83}. Using the Euler-Lagrange equation, we obtain the
equation of geodesics
\begin{eqnarray}\label{eq:geo}
\dot{t}^2-\frac{\zeta^2}{a^2}-2\epsilon=0\;,
\end{eqnarray}
where $\epsilon=0$ for the null geodesics and $\epsilon=1/2$ for
the time-like geodesics, and $\zeta$ is an integration constant
that represents the momentum of the particle. Therefore,
$\tau$ is integrated as
\begin{eqnarray}
\tau=+\int_0^{+\infty}{\frac{1}{\sqrt{2\epsilon+\frac{\zeta^2}{a^2}}}}dt=\infty\;,
\label{eq:geo1}
\end{eqnarray}
for the future-pointing geodesics, and
\begin{eqnarray}
\tau=-\int_0^{-\infty}{\frac{1}{\sqrt{2\epsilon+\frac{\zeta^2}{a^2}}}}dt=\infty\;,
\label{eq:geo2}
\end{eqnarray}
for the past-pointing geodesics, respectively.
Equations~(\ref{eq:geo1}) and (\ref{eq:geo2}) show that the space-time
is complete for both null and time-like geodesics.

\subsection{horizons}

In general, there are three horizons in the Friedmann-Robertson-Walker
space-time, i.e., the particle horizon, event horizon and apparent
horizon. In a universe with a finite age, a photon would only
propagate a finite distance in that finite time and the volume of
space from which we can receive information at a given time is also
limited.  The boundary of this volume is called the particle horizon
which is given by
\begin{eqnarray}
d_p=a\int_{t_i}^{t}{\frac{dt}{a}}\;,
\end{eqnarray}
where $t_i$ is the beginning of the universe. For the cyclic
universe presented above, we have $t_i=-\infty$ and $a_{min}\leq
a\leq a_{max}$. So the corresponding particle horizon is infinite.

The event horizon is defined as the complement of the particle
horizon. It encloses the set of points from which photons sent at a
given time $t$ will never be received by an observer in the
future. For our cyclic universe, we find the physical size of the
event horizon is also infinite,
\begin{eqnarray}
d_e=a\int_{t}^{+\infty}{\frac{dt}{a}}=\infty\;.
\end{eqnarray}

The apparent horizon is a marginally trapped surface with vanishing
expansion and has been argued to be a causal horizon for a dynamical
space-time. The apparent horizon is associated with the Hawking
temperature, gravitational entropy and other thermodynamical aspects
\cite{hay:99,bak:00,cai:05}. The first law of thermodynamics for the
apparent horizon has been derived not only in general relativity but
also in several alternative theories of gravity, including the Lovelock,
nonlinear, scalar-tensor, and braneworld theories
\cite{gong:07,cai:070,cai:071,cai:072,cai:073,she:070,she:071}.
Therefore, it is of great importance for us to consider the apparent
horizon in the cyclic model of the universe. For the spatially flat universe,
the physical size of the apparent horizon is given by
\begin{eqnarray}
d_a=\frac{1}{{H}}\;.
\end{eqnarray}
We recognize that it is exactly the Hubble scale or the Hubble
horizon. For our cyclic universe, we find there is a minimum of
the physical size of the apparent horizon, namely, the Planck
length. At the turning points from expansion to contraction (and
from contraction to expansion), the Hubble radius is infinite.
Since the particle horizon and event horizon are infinite, the
cosmic horizon problem is solved.
\section{entropy of the cyclic universe and violation of second law of thermodynamics}\label{sec:entr}
The cyclic universe constructed above is precisely cyclic in the
evolution of the scale factor of the universe. The early work of
Tolman \cite{tolman:1930s} in the 1930s showed that such a
universe is in fact impossible as long as the usual laws of
thermodynamics hold. The second law implies that the entropy
always increases, and this implies that from one cycle to the
next, the universe must grow. We find that the second law of
thermodynamics is actually violated in the presence of phantom
matter. So Tolman's proof does not apply. In the next, we shall
show this point in detail.

Many researches \cite{entropy} on the relation between
thermodynamics and gravitation theories indicate that the entropy
of the universe is proportional to the apparent horizon area of
the universe, which is also implied by the holographic principle
\begin{eqnarray}
S=\frac{1}{4}A=\frac{1}{4}\cdot4\pi d_a^2=\frac{\pi}{H^2}\;,
\end{eqnarray}
where $d_a$ is the radius of cosmic apparent horizon defined by
Eq.~(22). Then the variation of cosmic entropy with respect to the
cosmic time $t$ is

\begin{eqnarray}\label{eq:entropy}
\frac{dS}{dt}&=&-\frac{2\pi}{H^3}\dot{H}\nonumber\\
&=&{3\pi}{d_a}\left(1+w{'}\right)\;.
\end{eqnarray}
Here $w{'}$ is the equation of state defined as $w'=\frac{\sum_i
p_i}{\sum_i \rho_i}$, where $i$ represents the $i$-th component of
matters. In the last step, we have used the Einstein equation,
$\dot{H}=-4\pi\left(\rho+p\right)$. Eq.~(\ref{eq:entropy}) reveals
that for the dust matter ($w'=0$), relativistic matter ($w'=1/3$)
and quintessence matter ($-1\leq w'\leq 1$) dominated expanding
universe, the entropy of the universe is always increasing.
However, for a phantom dominated expanding universe ($ w'<-1 $),
the entropy of the universe is decreasing. This is very much
similar to the finding of Babichev et al \cite{babi:04} that the
black hole mass (namely, the black hole entropy, $S\propto M^2$)
decreases with the accretion of phantom matter. Actually,
Ref.~\cite{ger:05} has noted that the entropy of the phantom
matter is negative. Then we conclude that the second law of
thermodynamics in the phantom dominated universe is violated. This
is closely related to the violation of null energy condition
($\rho+p>0 $) by phantom matter.

\section{quantum stability}\label{sec:quant}

In this section, we shall consider the quantum stability of this
cyclic universe model. In the approximation of minisuperspace, the
wave function of the universe depends only on one freedom, the
scale factor $a$. The Hamiltonian of the universe is given by
\begin{eqnarray}
\mathscr{H}=-\frac{1}{3\pi a}\left[p_a^2+U\left(a\right)\right]\;,
\end{eqnarray}
where
\begin{eqnarray}
p_a=-\frac{3\pi}{2}a\dot{a}\;,
\end{eqnarray}
is the momentum conjugate to $a$ and the potential $U(a)$ is given
by
\begin{eqnarray}
U\left(a\right)=\frac{9\pi^2}{4}a^2\left(1-\frac{8\pi}{3}a^2\rho\right)\;,
\end{eqnarray}
where $\rho$ is total energy density of the universe. Then the
Hamiltonian constraint $\mathscr{H}=0$ gives exactly the Friedmann
equation, Eq.~(\ref{eq:f}).

In the theory of quantum cosmology, the universe is described by a
wave function $\Psi(a)$. The conjugate momentum $p_a$ becomes a
differential operator $-i\frac{d}{da}$ and the Hamiltonian
constraint is replaced by the Wheeler-DeWitt (WDW) equation
\cite{wdw:67}
\begin{eqnarray}
\mathscr{H}\Psi=0\;,
\end{eqnarray}
i.e.,
\begin{eqnarray}
\left(-\frac{d^2}{da^2}+U\right)\Psi=0\;.
\end{eqnarray}

\begin{figure}[h]
\begin{center}
\includegraphics[width=9cm]{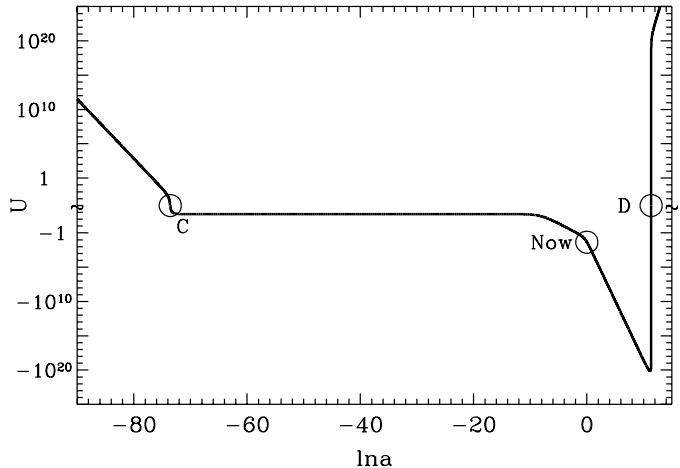}

\caption{The evolution of the potential $U(a)$. Points
$\textrm{C}$ and $\textrm{D}$ mark the minimum and maximum of the
scale factor $a$, respectively. The region between $\textrm{C}$
and $\textrm{D}$ is classically allowed. As the potential is
divergent when $a\rightarrow 0$ and $a\rightarrow \infty$, the
universe can not tunnel to the classically forbidden regions. }

\label{pot}
\end{center}
\end{figure}

Fig.~\ref{pot} shows the potential $U$ with respect to the scale
factor $a$. There is an oscillating region between the classical
turning points $\textrm{C}$ and $\textrm{D}$ which correspond to
the minimum and maximum of the scale factor, respectively. The
region between $\textrm{C}$ and $\textrm{D}$ is classically
allowed.

The semiclassical tunnelling probability as the universe bounces
at $\textrm{C}$ can be determined from

\begin{eqnarray}
\mathscr{P}\sim e^{-2S_{WKB}}\;,
\end{eqnarray}
where the tunnelling action are

\begin{eqnarray}
S_{WKB}=\int_{0}^{a_{min}}\sqrt{U}da=+\infty\;,
\end{eqnarray}
Since the tunnelling probability is vanishing, the cyclic universe
is able to last forever. In other words, this model of cyclic
universe is not immune to collapse and it is indeed quantum
stable.

\section{generation of scale-invariant power spectrum}\label{sec:scaleinvariant}

In the Cyclic Model \cite{paul:02} of the universe, {because the
equation of state $w'\gg1$ during the contraction phase, the
universe is homogeneous, isotropic and flat \cite{eri:04} with a
scale-invariant spectrum of density perturbations
\cite{gra:04,boy:06}. The $w'\gg1$ condition also ensures that
anisotropy is small and first order perturbation theory remains
valid until just before the bounce \cite{eri:04,paul:04}.

On the other hand, it is well-known that in order to generate
scale invariant curvature perturbations in a contracting universe,
two scalar fields are necessary. For the most recent models, see
Ref.~\cite{mingzhe:13}. Fig.~\ref{w} shows the evolution of $w'$
with the natural logarithm of the scale factor ($\textrm{ln} a$).
As shown in Fig.~\ref{w}, $w'\gg 1$ in the vicinity of bouncing
point $\textbf{B}$ (or $\textbf{D}$) (see Fig.~(1), Fig.~(2) or
Fig.~(3)). Thus we conclude that the flatness of spacetime and the
scale-invariant primordial power spectrum can also be generated in
the present cyclic universe model. Furthermore, we find the
scale-invariant power spectrum could be generated in the vicinity
of bouncing point $\textbf{C}$ (or $\textbf{E}$). In this section,
we shall study the generation of scale-invariant power spectrum in
the vicinity of bouncing points.

\begin{figure}[h] \begin{center} \includegraphics[width=9cm]{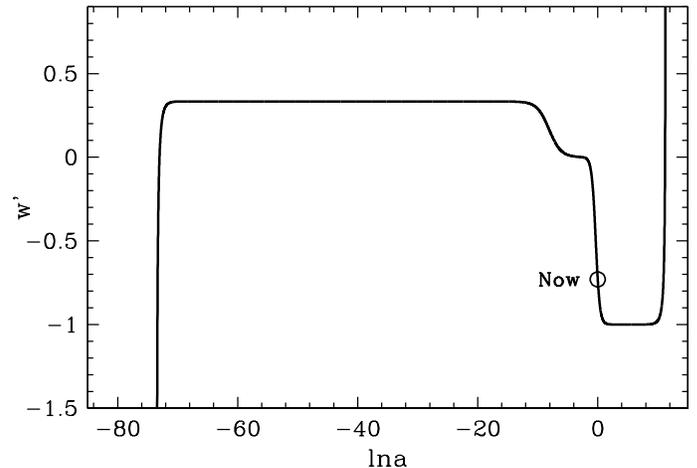}
\caption{Evolution of the equation of state $w'$ with the natural
logarithm of scale factor $\textrm{ln} a$. We have $\mid
w'\mid\gg1$ in the vicinity of the turning point from expansion to
contraction (and from contraction to expansion)} \label{w}
\end{center}
\end{figure}

\subsection{Bouncing point $\textbf{C}$ (or $\textbf{E}$)}
Around the bouncing point $\textbf{C}$ (or $\textbf{E}$, see
Fig.~(1), Fig.~(2) or Fig.~(3)), the evolution of the universe is
governed by the following Friedmann equation
\begin{eqnarray}\label{eq:rf}
3H^2&=&8\pi\left(-\frac{\rho_{\phi_{0}}}{a^6}+\frac{\rho_{r_{0}}}{a^4}\right)\;.
\end{eqnarray}
Rewrite above equation of motion with the conformal time $\eta$
and rescale both the cosmic comoving time and the scale factor, we
derive the evolution of the scale factor
\begin{eqnarray}\label{eq:ra}
a=\sqrt{1+\eta^2}\;.
\end{eqnarray}
Here the minimum of the scale factor is rescaled to be unit one
and $\eta=0$ denotes the bouncing moment. In the vicinity of
bouncing point, namely, $|\eta|\ll1$, the comoving Hubble
parameter $\mathscr{H}$ could be safely approximate as
\begin{eqnarray}\label{eq:rh}
\mathscr{H}=\frac{\eta}{{1+\eta^2}}\simeq \eta\;.
\end{eqnarray}
Fig.~(5) tells us the total equation of state in this epoch is
very much smaller than ${-1}$. So we can effectively use the
phantom matter to describe the matter content of the universe.
Because the evolution of the linear perturbations is determined by
the background evolution, namely, the evolution of the Hubble
parameter, we need not to know the expression of phantom potential
in great detail.

The equation of motion for the adiabatic scalar perturbations is
derived in Ref.~\cite{garriga:99}
\begin{eqnarray}\label{eq:reom}
v_{k}^{''}+\left(k^2-\frac{z^{''}}{z}\right)v_{k}=0\;,
\end{eqnarray}
where
\begin{eqnarray}\label{eq:rz}
z\equiv\frac{a}{H}\sqrt{\mid \rho+p \mid}\;,
\end{eqnarray}
and the prime denotes the derivative with respect to the cosmic
comoving time. Taking into account the Friedmann equation, the
acceleration equation and the expression of Hubble parameter,
Eq~(\ref{eq:rh}), we have
\begin{eqnarray}\label{eq:rzz}
z&\equiv&\frac{a}{H}\sqrt{\mid \rho+p
\mid}\propto\frac{a}{H}\sqrt{\mid\dot{H}\mid}\nonumber\\&=&\frac{a\left[\frac{1}{a}\left(\frac{\mathscr{H}}{a}\right)^{'}\right]^{1/2}}{\frac{\mathscr{H}}{a}}
\simeq\frac{1}{\eta}\;.
\end{eqnarray}
Here dot denotes the derivative with respect to cosmic time $t$
and $\mathscr{H}$ is the Hubble parameter in terms of the comoving
time $\eta$. So
\begin{eqnarray}\label{eq:reomf}
\frac{z^{''}}{z}=\frac{2}{\eta^2}\;.
\end{eqnarray}
Then the equation of motion for the adiabatic scalar perturbations
turns out to be
\begin{eqnarray}\label{eq:reomff}
v_{k}^{''}+\left(k^2-\frac{2}{\eta^2}\right)v_{k}=0\;.
\end{eqnarray}
Refs.~\cite{wands:99} and ~\cite{finelli:02} have shown that above
equation lead to exact scale invariant power spectrum.

\subsection{Bouncing point $\textbf{B}$ (or $\textbf{D}$)}
Similarly, around the bouncing point $\textbf{B}$ (or
$\textbf{D}$), the Friedmann equation is given by
\begin{eqnarray}\label{eq:rf}
3H^2&=&8\pi\left(\rho_{\Lambda}-{\rho_{\psi_{0}}}{a^2}\right)\;.
\end{eqnarray}
Rescale the cosmic comoving time and the scale factor, we obtain
the evolution of the scale factor
\begin{eqnarray}\label{eq:ra}
a=\frac{1}{\sqrt{1+\eta^2}}\;.
\end{eqnarray}
Here the maximum of the scale factor is rescaled to be unit one
and $\eta=0$ denotes the bouncing moment. So in the vicinity of
bouncing point $|\eta|\ll 1$, the comoving Hubble parameter
$\mathscr{H}$ is
\begin{eqnarray}\label{eq:rhh}
\mathscr{H}=-\frac{\eta}{{1+\eta^2}}\simeq -\eta\;.
\end{eqnarray}
The total equation of state in this epoch is very much larger than
unit one. Therefore we can effectively use the quintessence matter
(but negative scalar potential) to describe the matter content of
the universe while irrespective of the knowledge of the
quintessence potential.

We have
\begin{eqnarray}\label{eq:rzz}
z&\equiv&\frac{a}{H}\sqrt{\mid \rho+p
\mid}\propto\frac{a}{H}\sqrt{\mid\dot{H}\mid}\nonumber\\&=&\frac{a\left[\frac{1}{a}
\left(\frac{\mathscr{H}}{a}\right)^{'}\right]^{1/2}}{\frac{\mathscr{H}}{a}}\simeq\frac{1}{\eta}\;.
\end{eqnarray}
and
\begin{eqnarray}\label{eq:reomf}
\frac{z^{''}}{z}=\frac{2}{\eta^2}\;.
\end{eqnarray}
Then the equation of motion for the adiabatic scalar perturbations
is also give by

\begin{eqnarray}\label{eq:reomff}
v_{k}^{''}+\left(k^2-\frac{2}{\eta^2}\right)v_{k}=0\;.
\end{eqnarray}
In other words, the scale invariant power spectrum could be also
produced in the vicinity of bouncing point $\textbf{B}$ (or
$\textbf{D}$).

\section{conclusion and discussion}\label{sec:conclusion}

In conclusion, we have explored a cyclic model for the universe by
simply considering a phantom field and a quintessence field. In
every cycle of the cosmic evolution, the phantom field is dominant
in the very early universe and the quintessence field is dominant
in the far future.

There are infinite cycles of cosmic expansion and contraction in
this model. At the turning points from expansion to contraction or
from contraction to expansion, the cosmic energy density is zero.
So the Hubble horizon is infinite at these {turning points}. We
also find that the cosmic spacetime is geodesically complete for
both time-like and null geodesics. Therefore, there is no Big-bang
or Big-crunch singularity in the space-time. We calculate the
particle horizon and the event horizon, respectively, and find
they are infinite. So the horizon problem is vanishing. Using the
Wheeler-DeWitt equation, we show the system is quantum stable
although a negative scalar potential is present. Finally, we prove
that the flatness of spacetime and the primordial scale-invariant
power spectrum could be generated in this cyclic model which is
consistent with the findings of Gratton et al \cite{gra:04} and
Boyel et al. \cite{boy:06}.

\acknowledgments

We thank the referee for the perceptive and insightful report
which has improved the paper significantly. This work is supported
by the National Science Foundation of China under the Grant No.
10973014, No. 11373031 and the 973 Project (No. 2010CB833004).

\newcommand\ARNPS[3]{~Ann. Rev. Nucl. Part. Sci.{\bf ~#1}, #2~ (#3)}
\newcommand\AL[3]{~Astron. Lett.{\bf ~#1}, #2~ (#3)}
\newcommand\AP[3]{~Astropart. Phys.{\bf ~#1}, #2~ (#3)}
\newcommand\AJ[3]{~Astron. J.{\bf ~#1}, #2~(#3)}
\newcommand\APJ[3]{~Astrophys. J.{\bf ~#1}, #2~ (#3)}
\newcommand\APJL[3]{~Astrophys. J. Lett. {\bf ~#1}, L#2~(#3)}
\newcommand\APJS[3]{~Astrophys. J. Suppl. Ser.{\bf ~#1}, #2~(#3)}
\newcommand\JHEP[3]{~JHEP.{\bf ~#1}, #2~(#3)}
\newcommand\JCAP[3]{~JCAP. {\bf ~#1}, #2~ (#3)}
\newcommand\LRR[3]{~Living Rev. Relativity. {\bf ~#1}, #2~ (#3)}
\newcommand\MNRAS[3]{~Mon. Not. R. Astron. Soc.{\bf ~#1}, #2~(#3)}
\newcommand\MNRASL[3]{~Mon. Not. R. Astron. Soc.{\bf ~#1}, L#2~(#3)}
\newcommand\NPB[3]{~Nucl. Phys. B{\bf ~#1}, #2~(#3)}
\newcommand\CQG[3]{~Class. Quant. Grav.{\bf ~#1}, #2~(#3)}
\newcommand\PLB[3]{~Phys. Lett. B{\bf ~#1}, #2~(#3)}
\newcommand\PRL[3]{~Phys. Rev. Lett.{\bf ~#1}, #2~(#3)}
\newcommand\PR[3]{~Phys. Rep.{\bf ~#1}, #2~(#3)}
\newcommand\PRD[3]{~Phys. Rev. D{\bf ~#1}, #2~(#3)}
\newcommand\RMP[3]{~Rev. Mod. Phys.{\bf ~#1}, #2~(#3)}
\newcommand\SJNP[3]{~Sov. J. Nucl. Phys.{\bf ~#1}, #2~(#3)}
\newcommand\ZPC[3]{~Z. Phys. C{\bf ~#1}, #2~(#3)}
 \newcommand\IJGMP[3]{~Int. J. Geom. Meth. Mod. Phys.{\bf ~#1}, #2~(#3)}
  \newcommand\GRG[3]{~Gen. Rel. Grav.{\bf ~#1}, #2~(#3)}

\end{document}